\documentclass[aps,prd,twocolumn,nofootinbib,groupedaddress]{revtex4-1}

\usepackage{graphicx}
\usepackage{color}
\usepackage{lipsum}
\begin{document}

\title{Dilaton assisted two-field inflation from no-scale supergravity}

\author{Girish Kumar Chakravarty$^{1}$, Suratna Das$^2$, Gaetano Lambiase$^{3,4}$ and Subhendra Mohanty$^{1}$}

\affiliation{$^1$ Physical Research Laboratory, Ahmedabad 380009, India \\
$^2$ Indian Institute of Technology, Kanpur 208016, India \\
$^3$ Dipartimento di Fisica "E.R. Caianiello" Universit\'a di Salerno, I-84084 Fisciano (Sa), Italy \\
$^4$ INFN - Gruppo Collegato di Salerno, Italy}


%
\def\be{\begin{equation}}
\def\ee{\end{equation}}
\def\al{\alpha}
\def\bea{\begin{eqnarray}}
\def\eea{\end{eqnarray}}
\def\beas{\begin{eqnarray*}}
\def\eeas{\end{eqnarray*}}

\begin{abstract}
We present a two-field inflation model where inflaton field
has a non-canonical kinetic term due to the presence of a dilaton
field. It is a two-parameter generalization of one-parameter Brans-Dicke gravity in the Einstein frame.
We show that in such an inflation model the quartic and
quadratic inflaton potentials, which are otherwise ruled
out by the present Planck-{\it Keck}/BICEP2 data, yield scalar spectral index and tensor-to-scalar ratio in accordance
with the present data. Such a model yield tensor-to-scalar ratio of the order of $10^{-2}$ which
is within the reach of $B-$mode experiments like {\it Keck}/BICEP3, CMBPol and thus can
be put to test in the near future. This model yields negligible non-Gaussianity and no isocurvature 
perturbations upto slow-roll approximation. Finally, we show that such a model can be realised 
in the realm of no-scale supergravity.
\end{abstract}


\maketitle

\section{Introduction}
Inflation \cite{guth}, a rapid
accelerated expanding phase of the universe in its very early stage of evolution, provides a
mechanism of generation of tiny density fluctuations over the
homogeneous and isotropic background, which later evolve into large
scale structures like galaxies and clusters of galaxies in the
universe. It also explains the observed nearly scale invariant
spectrum of these density fluctuations on superhorizon scales.
Parameters, like scalar amplitude, scalar spectral index and
tensor-to-scalar ratio, predicted by inflationary paradigm are being
very accurately determined by the high-precision CMBR observations,
like WMAP and Planck. The recent Planck-2015 
data predicts the scalar amplitude and the scalar
spectral index as $10^{10}\ln(\Delta_{\mathcal R}^{2}) = 3.089\pm
0.036$ and $n_{s}= 0.9666 \pm 0.0062$ respectively at ($68 \%$ CL,
PlanckTT+lowP) \cite{Ade:2015lrj,Planck:2015}.  
Recently BICEP2/Keck Array CMB polarization experiments
combined with Planck analysis of CMB polarization and temperature data have 
put a bound on tensor-to-scalar ratio as $r_{0.05} < 0.07 \,(95\%$ CL)
 \cite{bicep2keck2015,Ade:2015xua}. Planck also confirms that the primordial perturbations, 
 to a good approximation, are indeed Gaussian in nature by constraining the primordial Non-Gaussianity
 (NG) amplitude as $f_{\rm NL}\sim0$~\cite{Ade:2015ava}.
 

There is a plethora of inflationary models which explain the key
inflationary parameters like scalar amplitude, scalar spectral index
and tensor-to-scalar ratio as observed in the CMB measurements and 
also account for the smallness of the non-Gaussianity parameter simultaneously. But it
is well known that the simplest single field slow-roll inflation
models with quartic $\lambda\phi^4$ and quadratic $m_{\phi}^2 \phi^2$
potentials are ruled out from the observations as they produce large
tensor-to-scalar ratio $r \simeq 0.26$ and $r\simeq0.13$, respectively, 
for $60$ e-foldings~\cite{Ade:2015lrj}.
One novel way of making the quartic potential of inflaton viable is
through what is now known as the Higgs inflation scenario
\cite{Bezrukov}, where the inflaton field $\phi$ is non-minimally
coupled to the curvature scalar $R$ (coupling term looks like $\xi
\phi^2 R$). This gives rise to very small tensor-to-scalar ratio
$r\sim 0.003$ for $N\approx60$. Such a model however encounters the
problem of unitarity violation in Higgs-Higgs scattering via graviton exchange 
at Planck energy scales because of very large curvature
coupling $\xi\sim10^4$ required to obtain the observed CMB amplitude~\cite{Giudice:2010ka}. Starobinsky model of
inflation which is $R^2$ correction to Einstein gravity is mathematically
equivalent to Higgs inflation model with quartic potential and
therefore it produces the similar small tensor-to-scalar ratio $r\sim 0.003$~\cite{starobinsky1}. 
On the other hand, generalized non-minimally coupled models with coupling
term $\xi \phi^a R^b$ and quantum corrected quartic potential $\lambda
\phi^{4(1+\gamma)}$, which are equivalent to power law inflation model
$R + R^\beta$, are studied in \cite{girish} which produce large $r\sim0.2$, and thus are disfavored by
present status of the CMB data~\cite{bicep2keck2015}. 
In the context of upcoming experiments like Keck/BICEP3 and CMBPol, it would be of 
interest to come up with models which predict $r$ close to the current bound of $r<0.07$.
From the theoretical perspective it would be desirable to examine scenarios where 
the $\lambda\phi^4$ or $m_{\phi}^2 \phi^2$ potentials, which generically occur in particle 
physics, are compatible with CMB observations.

In this work we study a two-field inflationary model where the
inflaton field $\phi$ is assisted by a dilaton field $\sigma$ and has
a non-canonical kinetic term due to the presence of a dilaton
field. Supergravity theories which are low energy limit of string theory contains several scalar fields
which can be of cosmological interest. The action of the model can be generically written as
\bea
S &=&\frac12\int d^4 x\sqrt{-g}\left[R-\nabla^\mu\sigma\nabla_\mu\sigma-e^{-\gamma\sigma}\nabla^\mu \phi\nabla_\mu \phi\right.\nonumber\\
&&~~~~~~~~~~~~~~~~~~~~~~~~~~~~~~~~~~~~\left.-2e^{-\beta\sigma} V(\phi)\right],\label{action-sigma-einstein2}
\eea
where $\beta$ and 
$\gamma$ are arbitrary independent parameters. Brans-Dicke (BD) gravity in Einstein frame (EF) is a special case where $\beta=2\gamma$ 
\cite{Brans:1961sx,Starobinsky:1994mh,GarciaBellido:1995fz,DiMarco:2002eb,Gong:1998nf}.
However BD gravity predicts $r$ larger than the observed limit, therefore generalization of BD theory 
is necessary for application to inflation. In this paper we generalize the BD theory in EF to a
two-parameter scalar-tensor theory where we treat $\beta$ and $\gamma$ as two independent arbitrary
parameters. Addition of one extra parameter allows us to obtain viable inflation
with otherwise ruled out quadratic and quartic potentials as we can have tensor-to-scalar 
ratio $r$ in the range of interest for forthcoming experiments. 

To note, it was shown by Ellis et al.~\cite{Ellis:2013xoa} that the inflaton field accompanied by a
moduli field $T$, which appear in string theories and have a no-scale supergravity 
form, give a potential for inflation equivalent to the $R+R^2$ Starobinsky model, producing $r\sim10^{-3}$.
We show in this paper that the above mentioned two parameter 
scalar-tensor theory can be obtained from no-scale supergravity theories
~\cite{Cremmer:1983bf,Ellis:1983sf,Lahanas:1986uc} which now can produce $r$ much larger than $10^{-3}$
and thus are observationally falsifiable by future experiments. Also in contrast to 
the supergravity embedding of the Starobinsky model, studied in~\cite{Kallosh:2014qta,Hamaguchi:2014mza},
where the imaginary part of the superfield $T$ $(i.e.$ axion $\phi)$ decreases rapidly and its real 
part $(i.e.$ dilaton $\sigma)$ drives the inflation, we will see that in our model dilaton-axion
pair evolves sufficiently during inflation and the axion acts as the inflaton. 

\section{Description of the model}
 First we look at the background dynamics of our model. Starting from the action 
(\ref{action-sigma-einstein2}), the equations of motion of the fields $\phi$ and $\sigma$ and 
the Friedmann equations can be obtained as 
\begin{eqnarray}
&&\ddot{\sigma}+3H\dot{\sigma}+\frac{\gamma}{2}e^{-\gamma\sigma}\dot{\phi}^2-\beta e^{-\beta\sigma}V(\phi)=0, \label{fullsigmadot} \\
&&\ddot{\phi}+3H\dot{\phi}-\gamma\dot{\sigma}\dot{\phi}+e^{(\gamma-\beta)\sigma}V'(\phi)=0, \label{fullphiddot} \\
&& 3 H^2 =\frac12\dot\sigma^2+\frac12 e^{-\gamma\sigma}\dot{\phi}^2 +e^{-\beta\sigma}V(\phi), \label{fullHubble} \\
&&\dot{H}=-\frac{1}{2}\left(\dot\sigma^2+e^{-\gamma\sigma}\dot{\phi}^2\right), \label{fullHdot}
\end{eqnarray}
where an over dot represents derivatives w.r.t. time and prime denotes
derivative with respect to $\phi$.  In the slow-roll regime when both
the fields slow-roll, terms with double time derivatives
can be neglected and therefore the background equations reduce to
\begin{eqnarray}
\hspace{-.8cm}&&3H\dot{\sigma}=\beta e^{-\beta\sigma}V(\phi),~~~~~~~3H\dot{\phi}=-e^{(\gamma-\beta)\sigma}V'(\phi),\label{phidot}\\
\hspace{-.8cm}&&3H^2=e^{-\beta\sigma}V(\phi),~~~~~~~~~\dot{H}=-\frac{\dot\sigma^2+e^{-\gamma\sigma}\dot{\phi}^2}{2} \label{Hdot}.
\end{eqnarray}
Here the full potential $W(\sigma,\phi)\equiv e^{-\beta\sigma}V(\phi)$ can be regarded as the product of potentials 
of the individual fields,
$U(\sigma)\equiv e^{-\beta\sigma}$ and $V(\phi)$, and thus 
we define the slow-roll parameters for both the fields in a usual way (following \cite{GarciaBellido:1995fz}) :
\begin{eqnarray}
\epsilon_\phi&\equiv&\frac{1}{2}\left(\frac{V'(\phi)}{V(\phi)}\right)^2,~~~~~~~~~~~~
\eta_\phi \equiv \frac{V''(\phi)}{V(\phi)},\nonumber\\
\epsilon_\sigma&\equiv&\frac{1}{2}\left(\frac{U_{\sigma}}{U}\right)^2=\frac{\beta^2}{2},~~~~~~~~
\eta_\sigma \equiv \frac{U_{\sigma\sigma}}{U}=\beta^2, \label{slowparameters}
\end{eqnarray}
where $U_\sigma\equiv\partial U/\partial\sigma$. To ensure the smallness of the slow-roll 
parameters we demand that the Hubble slow-roll parameter $\epsilon_H\equiv-\frac{\dot H}{H^2}\ll1$ during inflation. We notice that 
\begin{eqnarray}
\epsilon_H=\epsilon_\sigma+e^{\gamma\sigma}\epsilon_\phi, 
\label{epsilonH}
\end{eqnarray}
which implies that $\epsilon_\sigma\ll1$ and $e^{\gamma\sigma}\epsilon_\phi\ll1$ during inflation. 
Again, taking a time-derivative of the second equation of (\ref{phidot}) to obtain $\dot H$, one 
obtains $\epsilon_H=\frac{\ddot\phi}{H\dot\phi}+e^{\gamma\sigma}\eta_\phi-(\gamma-\beta)\beta$, 
which implies that $e^{\gamma\sigma}\eta_\phi\ll1$ and $\gamma\ll\beta+\frac1\beta\sim \frac1\beta$
(as $\beta$ is to be taken smaller than unity). 
We would show later on that the dilaton field $\sigma$ evolve slower than the inflaton
field $\phi$ throughout the inflationary phase which would enable us to treat $\sigma$ as a background field.

Furthermore, we would require the initial field values to calculate the inflationary 
observables such as $n_s$, $r$ and $f_{\rm NL}$. From the first equation of (\ref{Hdot}),
we notice that
\begin{eqnarray}
\sigma&=&\sigma_0+\beta \ln\left(\frac{a}{a_0}\right), \\
\int d\phi\frac{V(\phi)}{V'(\phi)} &=& -\frac{e^{\gamma\sigma_0}}{\beta \gamma} \left[\left(\frac{a}{a_0}
\right)^{\beta\gamma}-1\right], \label{phi-evo}
\label{sigma}
\end{eqnarray}
where subscript $0$ indicates the values of the quantities 60 e-foldings prior to end of inflation.
Defining $f(\phi)\equiv \int d\phi\frac{V(\phi)}{V'(\phi)}$ and requiring 
$\frac{a_f}{a_0}\gtrsim e^{\Delta N}$ (subscript $f$ denoting the quantities at the end of inflation)
for sufficient inflation, one gets
\begin{eqnarray}
\frac{1}{\beta\gamma}\ln\left[1+ \beta\gamma e^{-\gamma\sigma_0}(f(\phi_0)-f(\phi_f))\right] \gtrsim \Delta N.
\label{efolds}
\end{eqnarray}


The perturbation analysis of such a model has been extensively discussed in 
\cite{Starobinsky:1994mh, GarciaBellido:1995fz,DiMarco:2002eb}, where the comoving curvature perturbation is defined as
\begin{eqnarray}
\hskip-0.3cm \mathcal{R} = \Phi-\frac{H}{\dot{H}}\left(\dot{\Phi}+H\Phi\right)=\Phi+H\frac{\dot{\sigma}\delta\sigma
+e^{-\gamma\sigma}\dot{\phi}\delta \phi}{\dot{\sigma}^2+e^{-\gamma\sigma}\dot{\phi}^2},
\label{comov-curv-pert}
\end{eqnarray}
where $\Phi$ is the scalar metric perturbation in the longitudinal gauge. As we are
dealing with a multi-field inflationay model, $\mathcal{R}$ is not a frozen quantity on superhorizon scales and
 its time evolution is given by
\bea
\dot\mathcal{R}=\frac{k^2}{a^2}\frac{H^2}{\dot H}\Phi + \mathcal{S},
\eea
where $\cal{S}$ represents isocurvature (or entropy) perturbations given by
\bea
\hskip-0.3cm\mathcal{S} = \frac{2H (\beta \dot\sigma \dot\phi^2
V(\phi) e^{-\gamma\sigma} + \dot\phi \dot\sigma^2 V'(\phi))}
{e^{\beta\sigma}(\dot\sigma^2 + e^{-\gamma\sigma} \dot\phi^2)^2} \left(\frac{\delta\sigma}{\dot\sigma}
-\frac{\delta\phi}{\dot\phi}\right). \label{iso-pert}
\eea

Under slow-roll approximation, one can solve for the scalar perturbations of the model,
$\delta\sigma$, $\delta\phi$ and $\Phi$, on superhorizon scales, which turn out to be \cite{Starobinsky:1994mh}
\begin{eqnarray}
\hskip-0.7cm\frac{\delta\sigma}{\dot\sigma}&=&\frac{c_1}{H}-\frac{c_3}{H},~~~~~~~~~~~
\frac{\delta \phi}{\dot \phi}=\frac{c_1}{H}+\frac{c_3}{H}
\left(e^{-\gamma\sigma}-1\right),\label{super-horizon-eqns} \\
\hskip-0.7cm\Phi&=&-c_1\frac{\dot H}{H^2}+c_3\left[\frac{1}{2} \left(\frac{V'(\phi)}
{V(\phi)}\right)^2 \left(1-e^{\gamma\sigma}\right) -\frac{\beta^2}{2}\right],
\label{super-horizon-eqns3}
\end{eqnarray}
where $c_1$ and $c_3$ are the time independent integration constants and can be fixed using initial conditions. 
In the above expression (\ref{super-horizon-eqns3}), terms proportional to $c_1$ represent the adiabatic modes
while those proportional to $c_3$ represent the isocurvature modes. 
Using eq.s~(\ref{phidot})-(\ref{slowparameters}), the comoving curvature perturbation~(\ref{comov-curv-pert})
can be simplified to the form: 
${\mathcal R} \simeq\Phi + c_1-c_3+c_3 \epsilon_\phi(\epsilon_\sigma + e^{\gamma\sigma} \epsilon_\phi).$ 

Since from eq.(\ref{super-horizon-eqns3}), it is clear that all the terms in $\Phi$ are proportional to
$(c_1,c_3)\times$ slow-roll parameters, therefore we will ignore the potential $\Phi$ compared to 
$c_1$ and $c_3$ in ${\mathcal R}$. From eq.~(\ref{super-horizon-eqns}), 
we can calculate $c_1$ and $c_3$. Substituting $c_1$ and $c_3$, 
the comoving curvature perturbation on super horizon scales becomes
\begin{eqnarray}
{\mathcal R} =H \frac{\delta \phi}{\dot \phi} e^{\gamma\sigma} A + H \frac{\delta\sigma}{\dot\sigma} B,
\label{comov-curv}
\end{eqnarray}
where $A=\epsilon_\phi/\epsilon_H$ and $B=\epsilon_\sigma/\epsilon_H$. 

Now we look at the observables predicted by this model. The mode functions for
superhorizon fluctuations of $\sigma$ and $\phi$, evaluated
at horizon crossing $k = a(t_k) H(t_k)$, are $\langle|\delta\sigma(k)|^{2}\rangle =H^{2}(t_k)/2k^3$
and $\langle|\delta\phi(k)|^{2}\rangle =e^{\gamma\sigma(t_k)} H^{2}(t_k)/2k^3$.
Therefore, the power spectrum of comoving curvature perturbations becomes
\begin{eqnarray}
{\mathcal P}_{\mathcal R}\equiv\frac{k^3}{2\pi^2}\langle{\mathcal R}^2\rangle 
=\frac{H^2}{8\pi^2\epsilon_H}. \label{powerR1}
\end{eqnarray}

 The tensor power spectrum retains its generic form, because the action involves only minimal Einstein curvature term $R$, given by
$\mathcal{P}_{\mathcal T}=8H^2(t_k)/4\pi^2$,
which yields the tensor-to-scalar ratio as 
\begin{eqnarray}
r\equiv\frac{\mathcal{P}_{\mathcal T}}{\mathcal{P}_{\mathcal R}}=16\epsilon_H.
\label{tensor}
\end{eqnarray}
The scalar spectral index $n_s$ in this model is
\bea
\hskip-0.3cm n_s-1\simeq A \left[(2\eta_\phi-6\epsilon_\phi)e^{2\gamma\sigma} - \beta(2\beta+\gamma)e^{\gamma\sigma}\right]
-B \beta^{2}.\label{ns}\nonumber\\
\eea
It can be noted that in the limit $\beta\rightarrow0$ and $\gamma\rightarrow0$ (i.e. $A=1$ and $B=0$)
the forms of power spectrum, tensor-to-scalar ratio and 
spectral index  reduce to their standard forms in the single field slow-roll inflation.
%

\begin{figure}[b!]
 \centering
\includegraphics[width= 8cm]{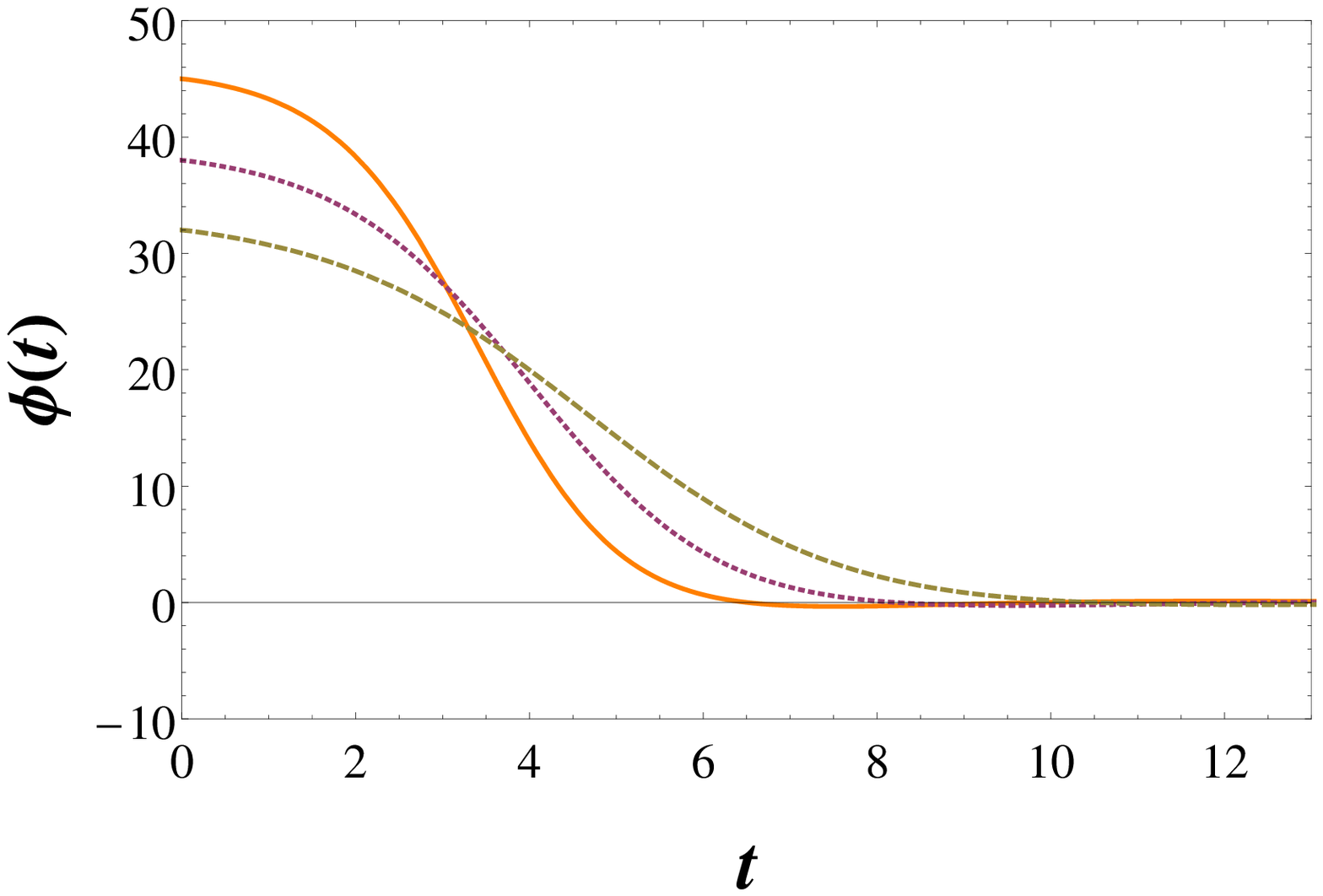}\vspace{0.3cm}
\includegraphics[width= 8cm]{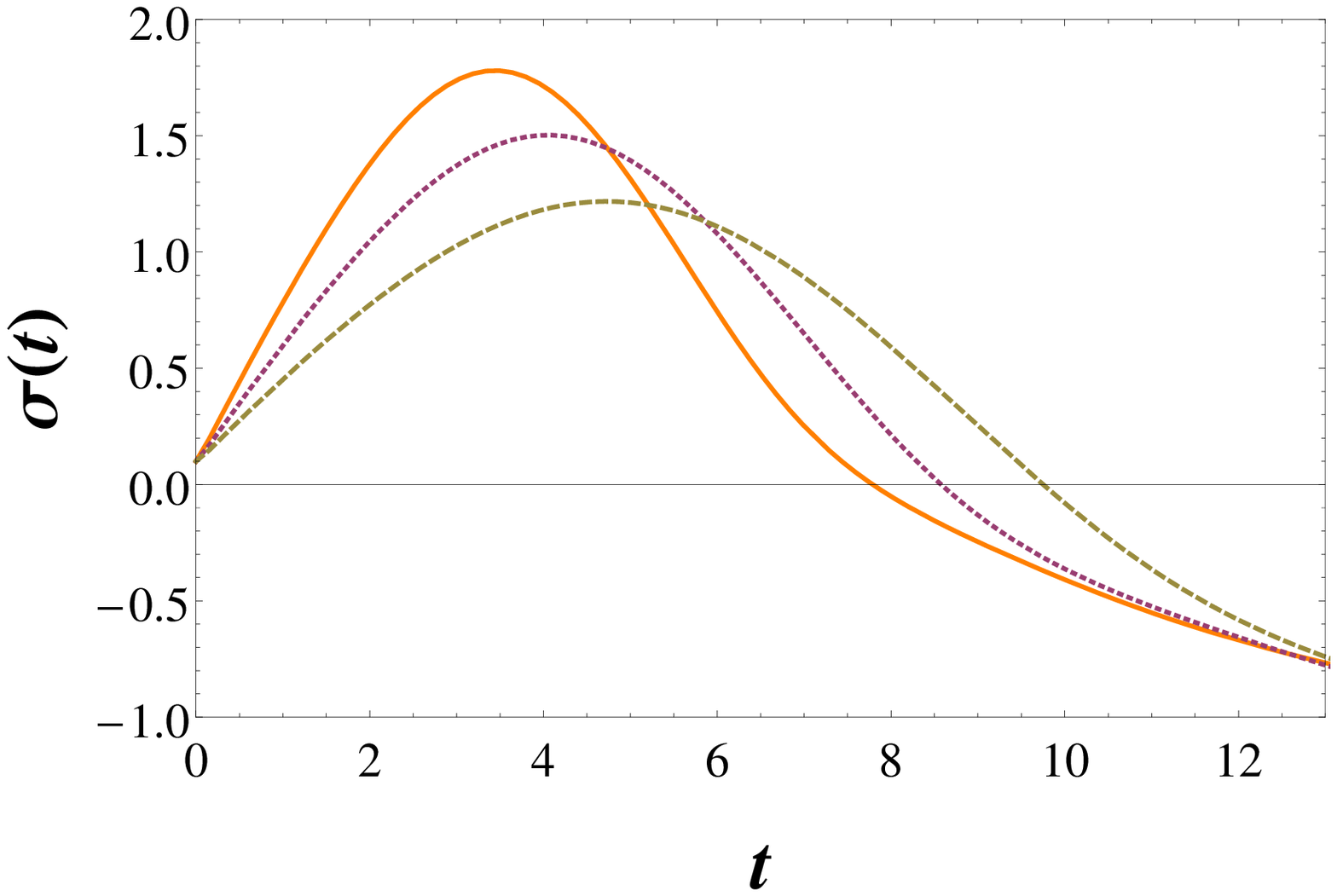}\hspace{0.0cm}
\caption{Evolution of the scalar fields w.r.t. time $t$ measured in the units of $m_{\phi}^{-1}$ is shown.}
 \label{fig1}
\end{figure}
\begin{figure}[t!]
 \centering
\includegraphics[width= 8cm]{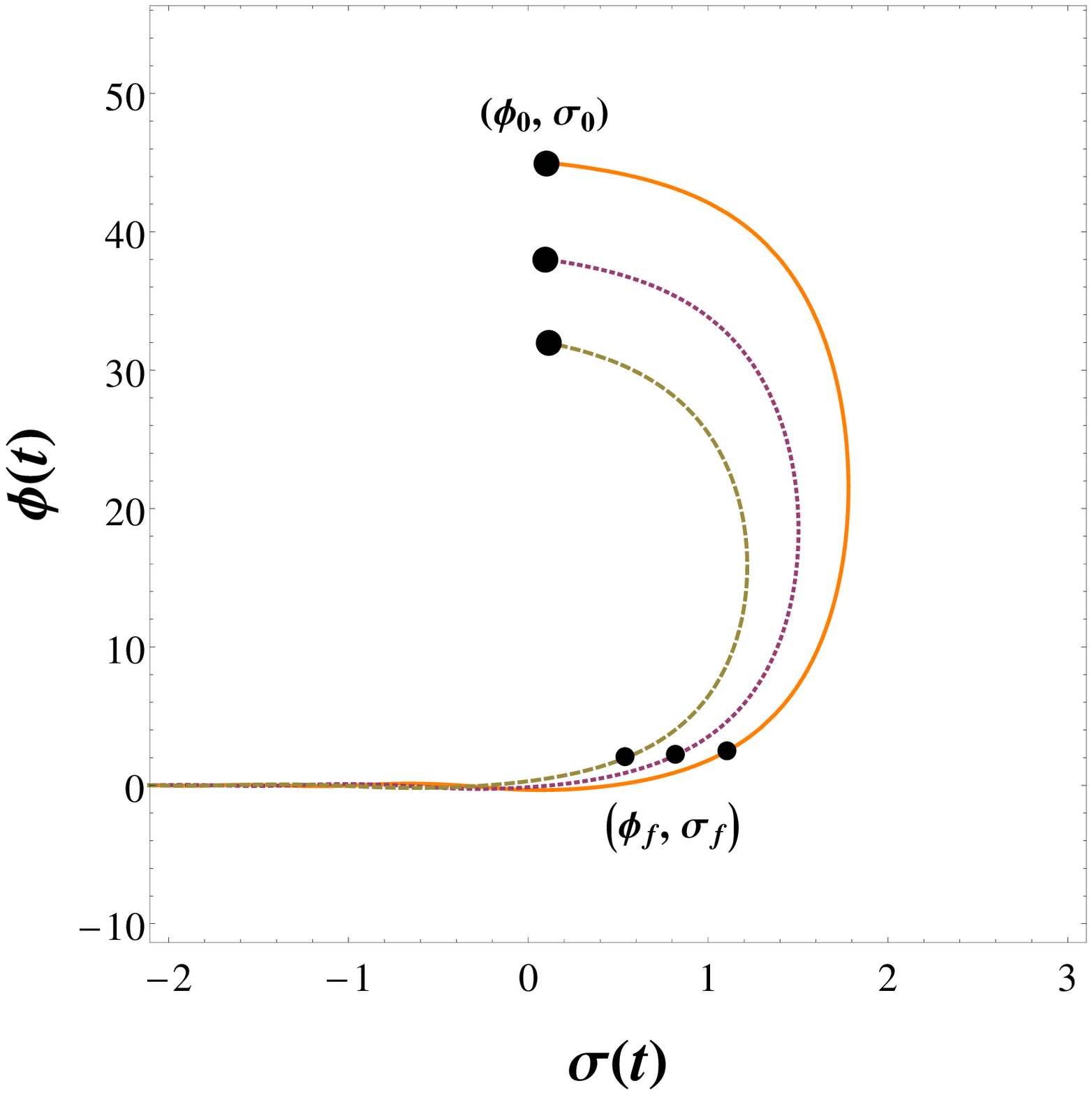}\vspace{0.3cm}
\includegraphics[width= 8cm]{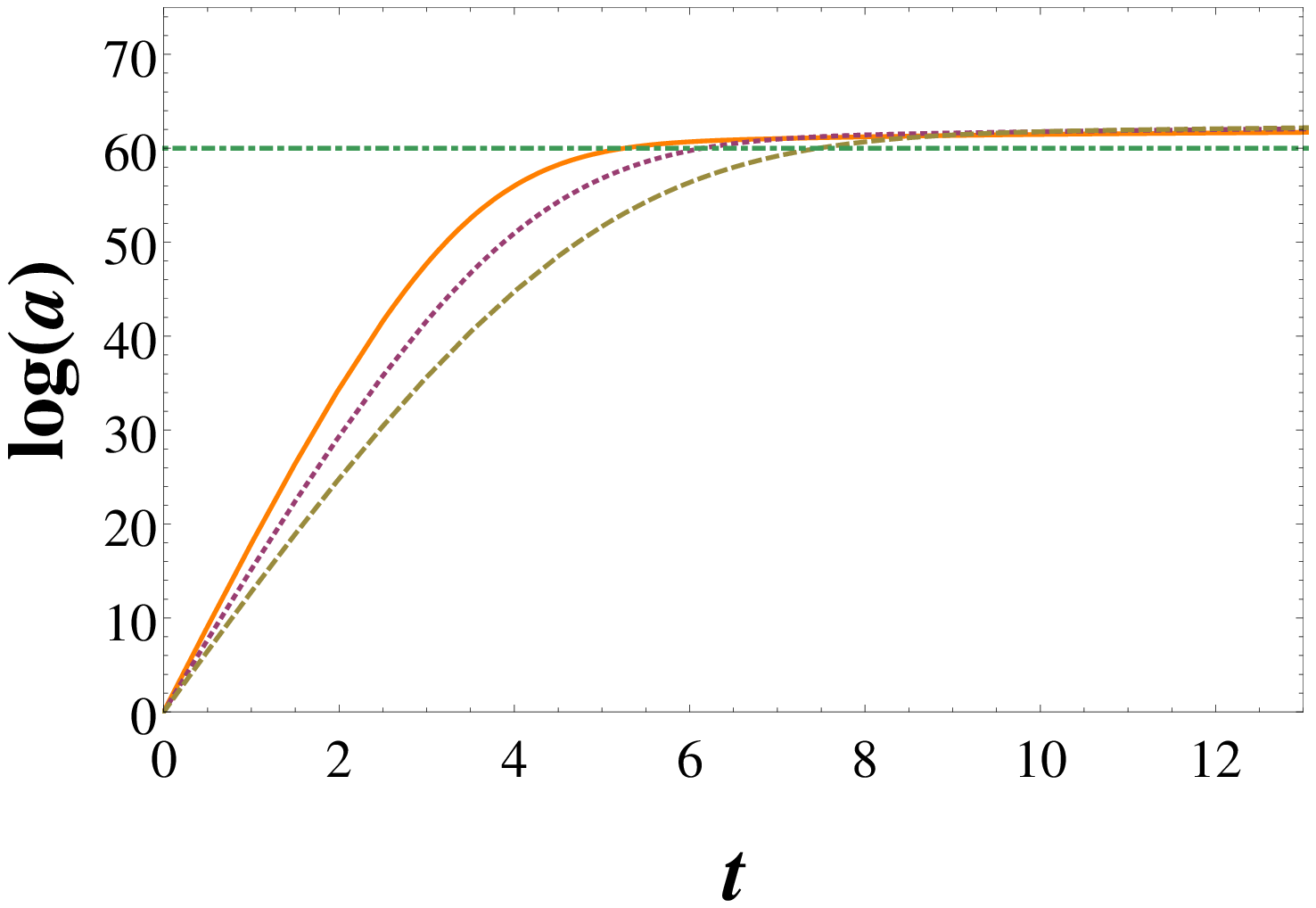}\hspace{0.0cm}
\caption{{\it Upper panel : } Evolution of the inflaton $\phi(t)$ with respect to dilaton $\sigma(t)$
is shown. Bigger black dots correspond to the field values when inflation starts and smaller dots correspond to the
field values at the end of inflation. We see that during inflation the evolution of dilaton
is much slower compared to inflaton. {\it Lower panel : } The cosmological evolution of the
scale factor is shown. The straight dot-dashed line represent the $\Delta N=60$ line where inflation ends.}
 \label{fig2}
\end{figure}

It is important to note that in our model the amplitude of the isocurvature perturbations, 
\begin{eqnarray}
\hskip-0.2cm{\mathcal P}_{\mathcal S} = \frac{k^3}{2\pi^2}\langle{\mathcal S}^2\rangle
= \frac{H^4}{\pi^2} \frac{\left[\beta e^{-\gamma\sigma}
\dot\phi V(\phi) + \dot\sigma V'(\phi)\right]^2}{e^{(2\beta-\gamma)\sigma}(\dot\sigma^{2}
+ e^{-\gamma\sigma} \dot\phi^2)^3}, \label{powerS}
\end{eqnarray}
vanishes in the slow-roll approximation (using (\ref{phidot})). 
The multifield models with non-canonical kinetic term posses a strong single-field attractor 
solution~\cite{Kaiser:2013sna,Schutz:2013fua}
as has also been observed in this case. But generally these multifield models produce isocurvature perturbations
which can also account for a large angular scale suppression of the power spectrum. But our case differs from such
multifield models as it does not produce any isocurvature perturbations upto slow-roll approximation.
%

\section{Analysis of the model} 

In our model, we would treat the $\phi$ field as the inflaton field which is assisted by 
the dilaton field $\sigma$ during the inflationary evolution. This can only be ensured if we confirm 
that the $\sigma$ field evolves slower than the $\phi$ field during the entire inflationary epoch. To show this, we first
 study the background evolution of both the scalar fields $\phi$ and $\sigma$
  by numerically solving the field eq.s (\ref{fullsigmadot}-\ref{fullHdot}).  We first
  treat the case when the inflaton field has quadratic potential $V(\phi)=m_{\phi}^{2}\phi^{2}/2$.
As some representative initial conditions, we choose $\sigma_0=0.1$ and $\phi_0=45$ (Solid), 
$\phi_0=38$ (Dotted), $\phi_0=32$ (Dashed) corresponding to $\beta=0.04$, $\beta=0.035$,
$\beta=0.03$ respectively. Also we fix $\gamma=2\sqrt{2/3}$ for each case,
which is required for the SUGRA derivation of this model studied in the later part of this paper.
The initial conditions are chosen carefully such that we get correct $n_s$ and $r$ for $\Delta N\simeq60$ e-folds.
In FIG.~\ref{fig1}, we show the time evolution of the fields $\phi$ and $\sigma$ where time is given in the units 
of $m_{\phi}^{-1}$. The different colors in the figure correspond to different initial conditions as described above.
In FIG.~\ref{fig2} (upper panel) shows the evolution of the fields in $(\sigma,\phi)$ plane. 
This plot shows that during $60$-efolds inflation, dilaton $\sigma$ evolves much slower compared to inflaton 
$\phi$. After the end of inflation, inflaton goes to its minimum value  $\phi=0$.
 Such background evolution of the fields also ensure that the background spatial metric evolves 
(quasi-)exponentially during inflation which has been depicted in the lower panel of FIG.~\ref{fig2}.
Also we checked that for the case of quartic potential $V(\phi)=\lambda \phi^{4}/4$, the fields evolve in a similar way
ensuring that $\phi$ can be treated as an inflaton field.

We now analyze the observable parameters for inflaton potential $V(\phi)=\lambda_n \phi^{n}/n$. 
From eq.~(\ref{sigma}), we find $\sigma_f = \sigma_0 + \beta \Delta N$. 
Using $\epsilon_H=1$, which is the condition for the end of inflation, we obtain $\phi_f= n e^{\gamma \sigma_{f}/2}/\sqrt{2-\beta^2} $. 
From eq.(\ref{efolds}), the field value $\phi_0$ can be expressed in terms of $\phi_f$ and $\sigma_0$ as
$\phi_0^{2} \simeq \phi_f^{2}+ \frac{2n}{\beta\gamma}e^{\gamma\sigma_0}\left(e^{\beta\gamma \Delta N}-1\right).$
\begin{figure}[t!]
 \centering
\includegraphics[width= 8cm]{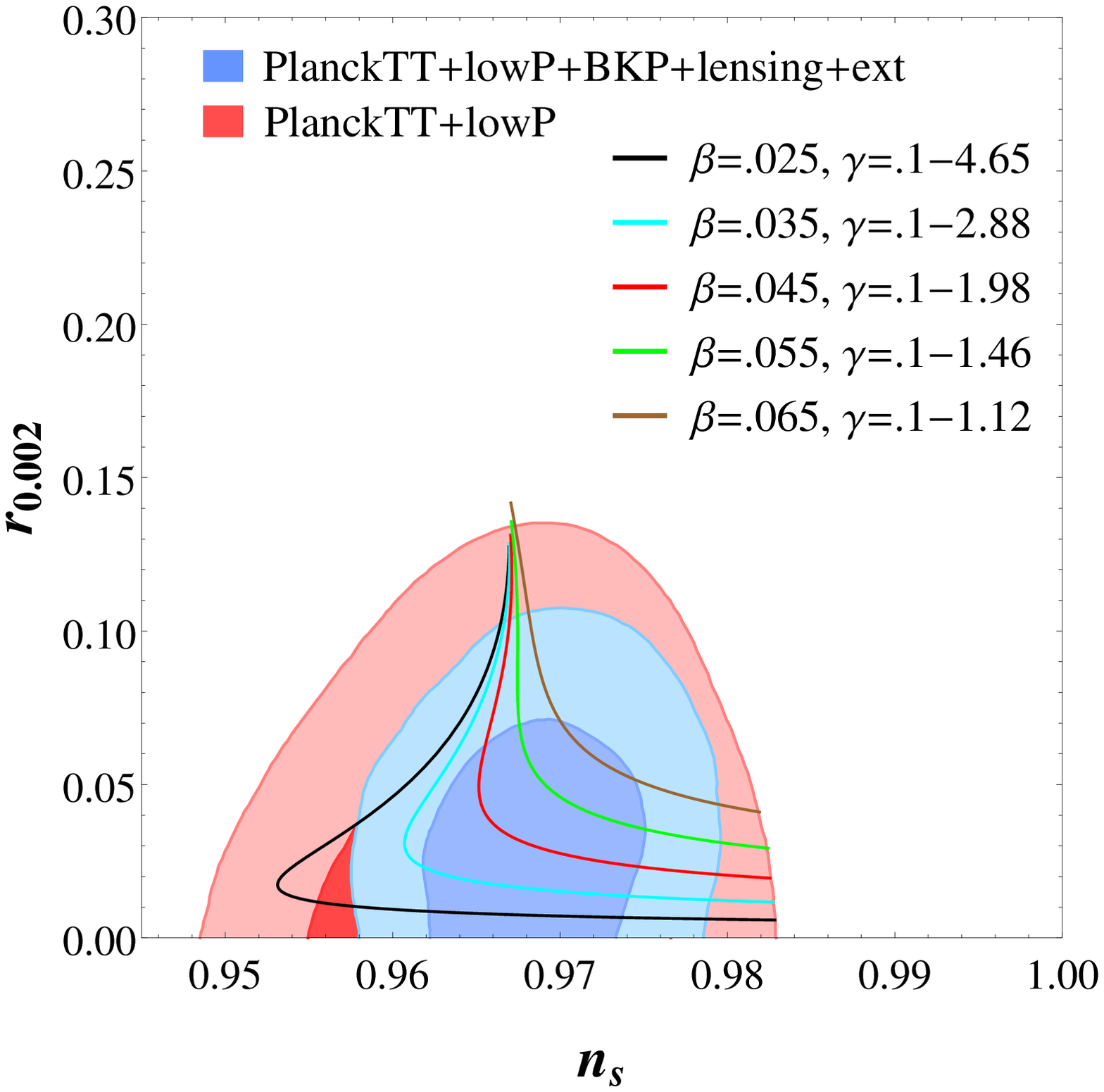}\vspace{0.3cm}
\includegraphics[width= 8cm]{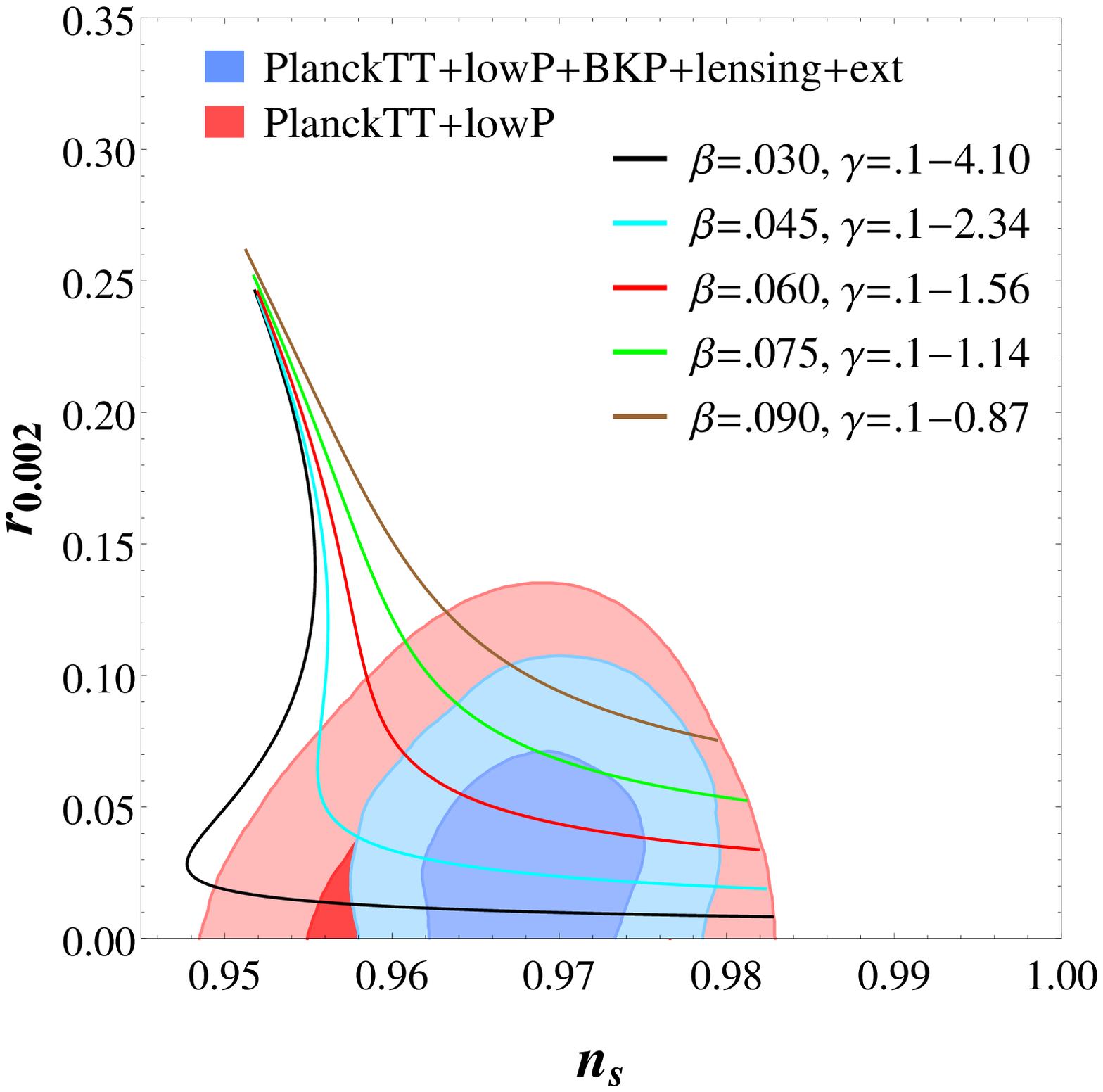}\hspace{0.0cm}
\caption{The $n_s - r$ predictions of the model for quadratic (upper panel) and quartic (lower panel)
potentials are shown with various contour lines and compared with $1\sigma$ and $2\sigma$ contours of 
the Planck observations \cite{Ade:2015xua}. We take $\Delta N = 60$ and $\sigma_0=0.1$. In both the figures
   the range of values of $\gamma$ increases along the curves from top to bottom. It is also manifest that as the
   values of $\beta$ and $\gamma$ goes to zero, $n_s$ and $r$ values converges to standard slow-roll inflation 
   predictions.}
 \label{fig3}
\end{figure}
Now we substitute $\phi_0$ into eq.s~(\ref{powerR1}), (\ref{tensor}) and (\ref{ns}) to
give $n_s$, $r$ and ${\mathcal P}_{\mathcal R}$ in terms of $\sigma_0$, $n$, $\Delta N$, $\beta$ and $\gamma$.
For $\Delta N =60$ e-folds and for the choice $\sigma_0 = 0.1$ with various choices of the parameters $\beta$
and $\gamma$, the $n_s-r$ predictions for quadratic $(n=2)$ and quartic $(n=4)$ potentials are shown in the 
FIG.~\ref{fig3}. 
For $\sigma_0=0.1$, $\Delta N =60$ and for the range of the parameter values 
of ($\beta,\gamma$) as shown in FIG.~\ref{fig3}, we find inflaton mass in the range $\lambda_2=m_{\phi}^{2}
\sim 10^{-11}-10^{-14}$ and self-coupling in the range $\lambda_4=\lambda\sim 10^{-13}-10^{-17}$. 
E.g. for the choice $\beta=0.05$, $\gamma=0.7$, which can produce $n_s\simeq 0.9666$, $r\simeq0.06$,
gives $m_{\phi} \approx 2\times10^{-6}$. And for $\beta=0.06$, $\gamma= 1$, which produces $n_s\simeq 0.964$,
$r\simeq0.05$, gives $\lambda\approx 10^{-16}$. Therefore, in this model with quadratic and
quartic potentials, similar to the case of single-field slow-roll inflation, we require light inflaton mass
and fine-tuning of the inflaton self-coupling in order to fit the observed CMB amplitude. However, unlike the Higgs 
inflation which predicts very small $r\approx0.003$ and standard single-field inlation with quadratic and quartic potentials give
large $r$, the two field model can give $r$ close to the present bound $r<0.07$. For the above mentioned initial conditions, 
coupling constants and parameter values, the running of the spectral index
$\alpha_s \equiv \frac{dn_s}{d\ln k}\simeq \frac{1}{H}\frac{dn_s}{dt}$ comes out to be 
$\alpha_s\simeq -6.7\times 10^{-4}$ and $\alpha_s\simeq -1.2\times 10^{-3}$ for quadratic and quartic potentials, respectively, fully
consistant with the Planck observation $\alpha_s =−0.0084 \pm 0.0082 ~(68 \% CL, Planck TT+lowP)$~\cite{Ade:2015lrj}.

Besides yielding $r$, $n_s$ and $\alpha_s$ within the observational bounds, a viable inflationary model should not produce
large Non-Gaussianity (NG) to remain in accordance with observations. NG in multifield
models where the fields have non-canonical kinetic terms has been calculated in Ref.s~\cite{Seery:2005gb,Choi:2007su}.
Following these Ref.s, we calculated the non-linearity parameter $f_{NL}$ which characterizes the amplitude of NG.
We find that for the range of parameters values as shown in FIG.~\ref{fig1},
$f_{\rm{NL}}\sim \mathcal{O}(10^{-2})-\mathcal{O}(10^{-3})$ consistent with the observations. 
Also we find that $f_{\rm{NL}}$ does not depend on initial 
value of the dilaton $\sigma_0$ and coupling constants for the considered chaotic form of potential.

\section{No-scale SUGRA realisation of the model} 

In this section we would show that, such a two-field inflationary model
can be realised in the realm of no-scale Supergravity.
The two-field models of inflation with string motivated tree-level no-scale K\"ahler
potential in no-scale supergravity framework are
analyzed in \cite{Casas:1998qx,Ellis:2014gxa,Ellis:2014opa,Ferrara:2014ima}. 
The F-term scalar potential in EF is determined from K\"ahler function given in terms of K\"ahler potential 
$K(\phi_{i},\phi^{*}_{i})$ and superpotential $W(\phi_{i})$ as $G \equiv K + \ln W +\ln W^{\ast}$, 
where $\phi_{i}$ are the chiral superfields. In the supergravity action, potential and kinetic
terms in EF are given by
\be
\hskip-0.2cm V=e^{G}\left[\frac{\partial G}{\partial \phi^{i}} K^{i}_{j*} \frac{\partial G}{\partial \phi^{*}_{j}}-3 \right],~~~~~
\mathcal{L}_{K}=K_{i}^{j*} \partial_{\mu}\phi^{i} \partial^{\mu}\phi^{*}_{j}, \label{LVLK}
\ee
where $K^{i}_{j*}$ is the inverse of the K\"ahler metric $K_{i}^{j*} \equiv 
\partial^{2}K / \partial\phi^{i}\partial\phi^{*}_{j}$.
We consider the K\"ahler potential of the following form
\bea
K=-3 \ln[T+T^{\ast}]+ \frac{b \rho \rho^{\ast}}{(T+T^{\ast})^{\omega}},\label{KP}
\eea
here $T$ is the two component chiral superfield whose real part is the dilaton and imaginary part
is an axion. We identify axion as the inflaton of the model, and $\rho$ is an
additional matter field with modular weight $\omega$. In typical orbifold string compactifications with three
moduli fields, the modular weight $\omega$ has value $3$~\cite{Dixon:1989fj,Casas:1998qx,Ellis:2014gxa}.
 Here we shall treat $\omega$ as a
phenomenological parameter whose value can have small deviation from the canonical value $3$ which may 
be explained via string loop corrections to the effective supergravity action \cite{Derendinger:1991hq}.
In this model to obtain the correct CMB observables, the parameter $(3-\omega)$ has to be fine tuned
to the order of $10^{-2}$.

For the complete specification of supergravity, we assume the superpotential as
$W=\lambda_{m} ~\rho ~T^{m}\label{SP}$.
We can decompose $T$ field in its real and imaginary parts parametrized by
two real fields $\sigma$ and $\phi$, respectively, as
$T=e^{-\sqrt{2/3}\sigma} + i \sqrt{2/3}\phi.$
\begin{figure}[t!]
 \centering
\includegraphics[width= 8cm]{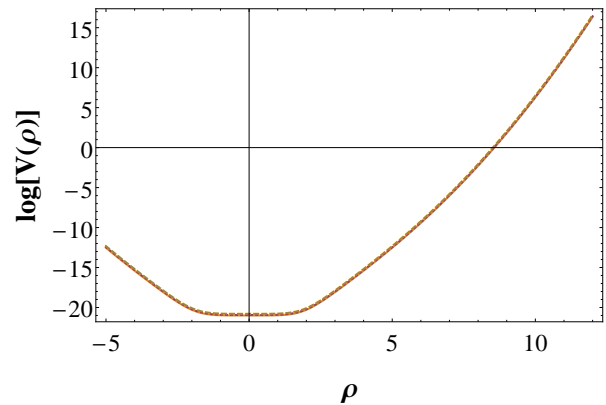}
\caption{Inflationary evolution of the potential $V(\rho)$ is shown for the three initial condition 
as discussed before. We see that during inflation the potential for the field $\rho$
is exponential steep and therefore the field $\rho$ rapidly falls towards the minima of the
potential and stabilizes at $\rho=0$.}
 \label{fig4}
\end{figure}

 The evolution of the matter field $\rho$ is constrained by the exponential factor $e^{K}$ via $e^{G}$ in 
the scalar potential (\ref{LVLK}) as $V\propto e^{\frac{b \rho\rho^{\ast}}{(T+T^{\ast})^{\omega}}}$.
Since $(T+T^{\ast})^{-\omega}=2 e^{\frac{2}{3}\omega \sigma}\gtrsim2$ for $\omega\approx3$ and 
$\sigma>0$ during inflation. Therefore field $\rho$, due to its exponentially steep 
potential, is rapidly driven to zero at the start of the inflation and stabilizes at $\rho=0$
~\cite{Ellis:2014opa}. In FIG.~\ref{fig4}, we show the stabilization of the field $\rho$ for 
different initial conditions as discussed before for $(2m=n=2)$. We also checked the evolution 
of $\rho$ for $(2m=n=4)$ and found that it stabilizes in the similar fashion.
Therefore for vanishing $\rho$, the scalar potential and kinetic term (\ref{LVLK}) takes the simple form
\be
V=\frac{\lambda_{m}^{2} T^{m}T^{\ast m}}{b (T+T^{\ast})^{3-\omega}}, ~~~~~~ \mathcal{L}_{K}=
\frac{3 \partial^{\mu}T \partial_{\mu}T^{\ast}}{(T+T^{\ast})^{2}},\label{VLK}
\ee
which upon using the decomposition of $T$ becomes
\bea
\mathcal{L}_{K}&=&\frac{1}{2} \partial^{\mu}\sigma \partial_{\mu}\sigma +\frac{1}{2} e^{-\gamma \sigma}
\partial^{\mu}\phi \partial_{\mu}\phi,\label{LK1} \\
V&=& \frac{2^{\omega-3}\lambda_{m}^{2}}{b}~ e^{-\beta\sigma} \left[e^{\gamma\sigma}+\frac{2}{3}\phi^{2} \right]^{m},\label{V1}
\eea
where $\gamma=2\sqrt{2/3}\simeq 1.633$ and $\beta=(3-\omega)\sqrt{2/3}$. Since during inflation, 
dilaton $\sigma$ evolves much slower compared to inflaton $\phi$, see FIG.~\ref{fig2}, therefore
$e^{\gamma\sigma}\ll \phi^{2}$ and hence the first term inside the bracket in (\ref{V1}) can be neglected compared to second term. 
\begin{figure}[t!]
 \centering
\includegraphics[width= 8cm]{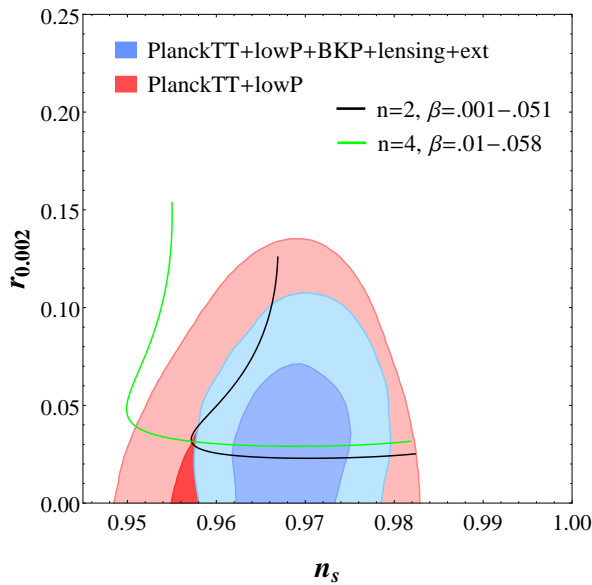}
\caption{The $n_{s}-r$ predictions for quadratic ($n=2$) and quartic ($n=4$) potentials
with a fixed value of $\gamma=2\sqrt{2/3}$ are shown and compared with 
$1\sigma$ and $2\sigma$ contours of the Planck observations \cite{Ade:2015xua}. 
The range of values of $\beta$ increases along the curves from top to bottom.}
 \label{fig5}
\end{figure}

Therefore, from (\ref{LK1}) and (\ref{V1}), the Lagrangian in EF becomes
\be
\mathcal{L}_{M}=\frac{1}{2} \partial^{\mu}\sigma \partial_{\mu}\sigma +\frac{1}{2} e^{-\gamma \sigma} 
\partial^{\mu}\phi \partial_{\mu}\phi + e^{-\beta\sigma} V(\phi),
\ee
where $V(\phi)=\lambda_{m}^{2}\phi^{2m}/2m$ and we set $b=2^{\omega}/6$ for quadratic
potential ($2m=n=2$) and $b=2\times2^{\omega}/9$ for quartic potential ($2m=n=4$). 
We see that the parameter $b$ is no new parameter and can be given in terms of $\omega$. 
For $\Delta N =60$ and $\sigma_0 = 0.1$, the $n_{s}-r$ predictions for a fixed value of
$\gamma=2\sqrt{2/3}$ and with varying $\beta$ are shown in FIG.~\ref{fig5}. 

\section{Conclusion}
To summarize, our two-field two-parameter inflationary model, where the inflaton field has
a non-canonical kinetic term due to the presence of the dilaton field, renders quartic and
quadratic potentials of the inflaton field viable with current observations. Unlike Higgs-inflationary scenario which predicts
very small tensor-to-scalar ratio $r\approx0.003$, 
this model can produce large $r$ in the range $r\sim10^{-1}-10^{-2}$
which would definitely be probed by future $B-$mode experiments
and thus such a model can be put to test with the future observations. 
Also this model produces no isocurvature perturbations upto 
slow-roll approximation and predicts negligible non-Gaussianity consistent with the observations.
In addition, we showed that this model can be obtained 
from a no-scale SUGRA model which makes this model of inflation phenomenologically interesting from the
particle physics perspective.
\vspace{0.3cm}
\begin{acknowledgments} 
Work of SD is supported by Department of Science and Technology, 
Government of India under the Grant Agreement number IFA13-PH-77 (INSPIRE Faculty Award). We thank 
the anonymous referees for their critical comments which improved the discussion of the model.
\end{acknowledgments}


\end{document}